\documentclass[twocolumn]{aastex7}
\usepackage{CJK}
\usepackage{amsmath}
\usepackage{graphicx}

\usepackage[utf8]{inputenc}

\usepackage{amsmath}
\renewcommand{\thesubsection}{\Roman{subsection}}
\usepackage{epstopdf}
\usepackage{commath}
\usepackage{natbib}
\usepackage{color}
\usepackage{xcolor}
\usepackage{lineno}
\usepackage{hyperref}
\usepackage{mathtools}
\usepackage[toc,page]{appendix}
\usepackage{lineno}
\usepackage{bm}
\usepackage[T1]{fontenc}
\usepackage[utf8]{inputenc}

  \renewcommand{\d}{\mathrm d}

\begin{document}

\title{Characterization of compressible fluctuations in solar wind streams dominated by balanced and imbalanced turbulence: Parker Solar Probe, Solar Orbiter and Wind observations}

\author[orcid=0000-0001-7063-2511]{C.A. González}
\email{carlos.gonzalez1@austin.utexas.edu}
\affiliation{Department of Physics, The University of Texas at Austin, Austin, TX 78712, USA}

\author[orcid=0009-0007-5716-5112]{C. Gonzalez}
\email{christiangon2004@utexas.edu}
\affiliation{Department of Physics, The University of Texas at Austin, Austin, TX 78712, USA}

\author[orcid=0000-0003-2880-6084]{A. Tenerani}
\email{Anna.Tenerani@austin.utexas.edu}
\affiliation{Department of Physics, The University of Texas at Austin, Austin, TX 78712, USA}
\correspondingauthor{C.A. González}
\email{carlos.gonzalez1@austin.utexas.edu}

\begin{abstract}
Characterizing compressible fluctuations in the solar wind is essential for understanding their role in solar wind acceleration and heating, yet their origin and evolution across different turbulence regimes remain poorly understood. In this study, we carry out a statistical analysis of the properties of compressible fluctuations in solar wind dominated by balanced and imbalanced turbulence. Using in-situ measurements from  Wind, Solar Orbiter and Parker Solar Probe, we investigate the scale dependence of density and magnetic pressure fluctuations and their correlations with plasma beta and radial distance. 
Our results indicate that solar wind compressibility is likely affected by both expansion effects and compressible dynamics governed by local plasma conditions. The non‑Alfvénic wind is dominated by anti-correlated fluctuations, whereas the Alfvénic wind contains a mixture of correlated and anti-correlated fluctuations, though the latter remain prevalent. 
While the anti-correlated component is consistent with MHD slow magnetosonic modes, the correlated (fast mode-like) component is not reproduced by predictions from either linear MHD theory or nonlinear models of forced compressible fluctuations. Nevertheless, the dominant slow mode component explains the observed dependence of $\delta B_\parallel/\delta B_\perp$ on $\beta$ and the enhanced density fluctuations measured by Parker Solar Probe.
This further suggests that slow mode waves contribute significantly to the compressible energy budget near the Sun and may play an important role in solar wind heating and acceleration close to the sun.

\end{abstract}

\section{Introduction}

The solar wind, the continuous plasma outflow from the Sun's corona into interplanetary space, is largely permeated by low-frequency Alfvénic turbulence, with strong correlations between plasma-velocity and magnetic-field fluctuations across a broad range of scales \citep{belcher1971, tu1995mhd, bruno2013solar}. In coronal-hole streams these Alfvénic fluctuations are predominantly outwardly propagating and carry enough energy to contribute significantly to plasma heating, although the dominant dissipation channels remain uncertain. A smaller population of sunward-propagating fluctuations is also present, and the imbalance between outward and inward wave populations tends to decrease with radial distance \citep{matthaeus1982measurement, bavassano2000evolution}. The relative fraction of counter-propagating fluctuations controls the strength of nonlinear interactions and thus the overall turbulent state \citep{chandran2008strong, schekochihin2022mhd}.

In addition to the Alfvénic component, the solar wind contains a smaller portion of compressible component at inertial and kinetic scales \citep{bavassano1989evidence, tu1994nature}. These fluctuations involve coupled variations in magnetic and thermal pressures; via density fluctuations they can generate Alfvén-speed gradients that enhance reflection and dissipation. Reflection-driven incompressible turbulence alone may be insufficient to heat the corona and accelerate the solar wind \citep{cranmer2023,verdini2019turbulent}, and compressible effects may supply additional reflection that strengthens the cascade and increases the heating rate \citep{asgari2021effects, verdini2019turbulent}. Compressibility can also trigger processes such as Alfv\'en-wave instabilities \citep{derby1978modulational} and nonlinear wave steepening \citep{cohen1974nonlinear}, which may contribute further to heating \citep{Gonzalez2021,gonzalez2023,gonzalez2024}.

A common characterization of compressible fluctuations uses the correlation between thermal- and magnetic-pressure variations (often via density fluctuations and fluctuations in magnetic-field magnitude). Under this diagnostic, fast-mode polarization corresponds to a positive correlation, whereas slow-mode polarization corresponds to a negative correlation. Observations near 1~au and beyond generally find negative correlations \citep{villante1982radial,roberts1987origin,bavassano2004compressive}, interpreted as pressure-balanced structures (PBS) and/or MHD/kinetic slow-mode fluctuations (including ion-acoustic and mirror modes) \citep{yao2011multi,howes2012slow,klein2012using,narita2015kinetic,verscharen2017kinetic}.

Closer to the Sun, the relative contributions of fast and slow modes remain uncertain. Helios observations between 0.3 and 1~au showed a mixture of fast and slow modes \citep{tu1994nature}. Early Parker Solar Probe observations reported both modes, with a larger contribution from the slow mode \citep{chaston2020mhd}, whereas other analyses found most of the power in the fast mode \citep{zhao2021analysis} and reported a reduction of the slow-mode component in the inner heliosphere \citep{chen2020}. More recent measurements in the inner heliosphere show enhanced density fluctuations associated with slow modes \citep{zhao2025transonic}, and remote-sensing observations of the outer corona indicate increased relative density fluctuations \citep{hahn2018density, miyamoto2014radial}. Consistently, 3D simulations of the inner heliosphere show the development of density fluctuations and the coexistence of fast and slow waves \citep{chiba2025density}.

Among the various mechanisms that can generate compressible fluctuations, those driven by large-amplitude Alfvénic fluctuations are particularly relevant in the Alfv\'enic solar wind. At sufficiently large amplitudes, nonlinearities enable parametric instabilities \citep{sagdeev1969nonlinear, derby1978modulational, goldstein1978instability, hollweg1994beat, buti2000hybrid, matteini2010kinetics}, while envelope-modulated Alfvén waves can drive density fluctuations through the plasma response to wave magnetic pressure \citep{cohen1974nonlinear}. Depending on parameters such as plasma beta and propagation angle, these processes can generate fast and/or slow modes \citep{hollweg1971density,cohen1974nonlinear,spangler1982properties,hada1993,mallet2021evolution}.

{Moreover, kinetic modes such as Kinetic Alfv\'en waves (KAW) and Kinetic Slow waves (KSW)\citep{narita2015kinetic}, could contribute to compressibility of the solar wind at sub-proton scales \citep{kiyani2013enhanced,roberts2018multi}. Plasma instabilities such as beam-driven instabilities \citep{gary1993theory} or those produced by temperature anisotropy \citep{klein2015predicted} can also generate compressible fluctuations, typically manifesting at kinetic scales.}

In this work, we present a statistical analysis of compressible fluctuations at inertial-range scales using Parker Solar Probe, Solar Orbiter, and Wind measurements, spanning heliocentric distances from 10~$R_\odot$ to 1~au. We investigate how the nature and level of compressibility evolve with distance and how they depend on fluctuation amplitude. While our primary focus is the Alfvénic, imbalanced wind, we also analyze intervals dominated by more balanced turbulence for comparison.

This paper is organized as follows. Section~\ref{datasec} describes the interval selection procedure, the datasets, and our methodology. Section~\ref{res} presents the main results, focusing on the scale dependence of magnetic and density fluctuations and their relationship with fluctuation's amplitude in balanced and imbalanced streams. We also examine the dependence on plasma beta and the correlation between density and magnetic pressure fluctuations. Section~\ref{conclusions} summarizes our conclusions and discusses broader implications.

\section{Data and methodology}
\label{datasec}

We investigated the properties of compressible fluctuations in the solar wind by using \textit{in-situ} measurements from three missions located in different regions of the heliosphere. At 1 au, we examined Wind spacecraft observations from 1999–2018, combining  downsampled 3-second resolution magnetic field data from the MFI instrument~\citep{Lepping1995} with ion measurements from the 3DP plasma instrument~\citep{lin1995}.{In addition, we used 92-second cadence proton density measurements from the Solar Wind Experiment (SWE)~\citep{ogilvie1995swe} as a reference for data quality of the density measurements in 3DP}. In the inner heliosphere, we used Parker Solar Probe (PSP) observations from 2019 to 2024 (Encounter 4th to 21st). We used magnetic field measurements from the FIELDS instrument suite~\citep{bale2016fields} and ion moments from the SPAN-I analyzer~\citep{Livi_2022}, along with electron density estimates from the quasi-thermal noise (QTN) signal measured by the Radio Frequency Spectrometer~\citep{moncuquet2020}. We used data within heliocentric distances less than 0.25~au to avoid low-quality ion moment data associated with field-of-view in SPAN-I instrument. All data is sampled at 3-second resolution.  Additionally, we included Solar Orbiter (SolO) data collected between 2022 and 2024, covering radial distances from 0.3 to 1~au. For SolO, we used 4-second resolution magnetic field measurements from the MAG instrument~\citep{horbury2020solar} and proton moments derived from the Solar Wind Analyzer–Proton and Alpha Sensor (SWA-PAS)~\citep{owen2020solar}.

In this study, we classified solar wind streams based on the normalized cross-helicity 
$$\sigma_c = ({ \langle | {\bf z}^{+} |^2 \rangle - \langle | {\bf z}^{-} |^2  \rangle})/({\langle | {\bf z}^{+} |^2  \rangle  + \langle | {\bf z}^{-} |^2  \rangle  }),$$
which quantifies the relative contributions of outward and inward Alfvén waves. The angle brackets $\langle ... \rangle$ indicate a moving average over a specified time interval, and the Els{\"a}sser variables are defined as ${\bf z}^{\pm} = \mathbf{\delta u} \pm \mathbf{\delta B}/\sqrt{(\mu_0 \rho)}$, where $\rho$ is the mass density. To compute $\sigma_c$, we used QTN data for PSP, whereas for Wind and SolO we employed proton density from 3DP and SWA, respectively. In this study, we define intervals with $|\sigma_c| < 0.25$ as non-Alfvénic (representing balanced turbulence), while intervals with $|\sigma_c| > 0.75$ are classified as Alfvénic wind (imbalanced turbulence). We do not distinguish between fast- and slow-Alfvénic wind since both display similar turbulence properties \citep{d2019slow,perrone2020highly,d2020large}.

{To examine the properties of large-scale inertial-range fluctuations, we first selected extended time intervals (5 days for Wind, 2 days for SolO, and 1 day for PSP), and identify sub-intervals where the normalized cross helicity ($\sigma_c$, computed at scales $\tau= \{8,4,2 \} \ \text{hr}$ for Wind, SolO, and PSP, respectively) stays roughly constant over time spans that are much longer than the characteristic fluctuation timescale. In this analysis, we used the outer scale of the turbulent fluctuations estimated from the autocorrelation function of the magnetic field fluctuations, 
$$
f(\tau) =\frac{\langle  \delta \mathbf{B}(t + \tau) \cdot \delta \mathbf{B}(t) \rangle_{\text{interval}}}{\langle | \delta \mathbf{B}^2 |\rangle},
$$
where $\delta \mathbf{B}(t) = \mathbf{B}(t) - \langle \mathbf{B} \rangle_{\text{interval}}$, and we define the outer scale as the time lag at which $f(\tau)$ decreases to $1/e$ of its initial value \citep{matthaeus1982stationarity}. 

Our selection criterion requires $\sigma_c$ to remain approximately constant for at least $30\tau_c$. This procedure results in a total of 801 intervals: Wind (249 Alfv\'enic; 56 non-Alfv\'enic), SolO (219; 67), and PSP (181; 29). The statistics of the resulting interval durations and their mean correlation time (computed for each interval) are as follows: for Alfv\'enic intervals, the average duration is $34.30 \pm 18.35$ hr with $\langle \tau_c \rangle = 0.44 \pm 0.35$ hr (Wind), $20.31 \pm 10.13$ hr with $\langle \tau_c \rangle = 0.49 \pm 0.34$ hr (SolO), and $10.03 \pm 7.25$ hr with $\langle \tau_c \rangle = 0.13 \pm 0.10$ hr (PSP). Non-Alfv\'enic intervals are, on average, shorter: $25.45 \pm 14.51$ hr with $\langle \tau_c \rangle = 0.91 \pm 0.57$ hr (Wind), $6.35 \pm 3.48$ hr with $\langle \tau_c \rangle = 0.61 \pm 0.38$ hr (SolO), and $3.66 \pm 2.42$ hr with $\langle \tau_c \rangle = 0.29 \pm 0.16$ hr (PSP).{The calculation of $\tau_c$ implicitly relies on both Taylor’s hypothesis and the assumption of stationarity, and earlier works have shown that estimates of $\tau_c$ are influenced by nonstationarity, the selected interval length as well as by solar wind conditions \citep{weygand2013magnetic,isaacs2015systematic,jagarlamudi2019inherentness}. Since the intervals length over which the correlation scale is calculated is long enough to include inertial range fluctuation, we used our calculated $\tau_c$ as a reference timescale to identify fluctuations in the inertial range (or, equivalently, the spatial scale $l_c = v_{sw} \tau_c$). On average the condition $\tau_c \gg \Omega_{ci}^{-1}$ and $l_c \gg \rho_i$ (and likewise for $d_i$) is satisfied by our intervals, with a typical scale separation of $10^{3}-10^{5}$.}

{We additionally generated a second dataset by dividing each long interval into sub-intervals of 2–3 hours. This procedure increases the number of samples while preserving inertial-range statistics. We removed all sub-intervals coinciding with Coronal Mass Ejections (CMEs) listed in the catalog of \citet{mostl2020prediction}.

{We note that \cite{howes2012slow} found that Wind intervals characterized by low density fluctuation amplitudes ($\delta n < 0.5 \mathrm{cm}^{-3}$) could be affected by instrumental noise. However, we do not find a similar behavior for $\delta n$ (see also Fig.~\ref{fig:append2}). To improve the quality of data for the Wind dataset, we followed a methodology similar to that described in \cite{bowen2018density}, excluding sub-intervals where the relative difference between the mean densities measured by 3DP and SWE is greater than 10\% ($|\langle n \rangle_{3DP} - \langle n \rangle_{SWE}|/\langle n \rangle_{SWE} > 0.1$).} 

For PSP, previous work has shown agreement within $20\%$ between the QTN and the proton SPAN-i density measurements \citep{martinovic2022plasma}. Accordingly, we excluded intervals for which the relative discrepancy between SPAN-i and QTN mean densities  over the interval exceeds 20\% ($|\langle n \rangle_{\mathrm{SPAN}}-\langle n \rangle_{\mathrm{QTN}}|/\langle n \rangle_{\mathrm{QTN}} > 0.2$), as well as intervals where QTN measured densities are below the instrument limit of $100\,\mathrm{cm}^{-3}$ \citep{moncuquet2020}. However the low-density criteria for PSP is relaxed for non-Alfvénic intervals. Close to the sun, PSP has sampled some of the most rarefied solar wind streams, often associated with heliospheric current sheet crossings \citep{cheng2024origin,phan2025parker}. This choice is motivated by the potential importance of these low-density periods, as well as by the limited number of non-Alfvénic intervals available in the PSP dataset.}
{After applying these filters to the short-interval dataset, we obtained 2867 Alfvénic sub-intervals for Wind, 2467 for SolO, and 460 for PSP; the non-Alfvénic set consists of 193 (Wind), 154 (SolO), and 30 (PSP) sub-intervals.}

We characterized the amplitude of fluctuations and compressibility using the following quantities: the normalized root-mean-square (rms) amplitude of magnetic fluctuations, ${\delta B}/{|\langle \mathbf{B}\rangle|} =[\langle|{\bf B - \langle B} \rangle|^2\rangle]^{1/2}/|\langle {\bf B}\rangle|$, 

the normalized rms of the magnetic field magnitude fluctuations (a proxy for magnetic compressibility) $\langle \delta|{\bf B}|^2\rangle^{1/2} /\langle |\mathbf{B}| \rangle={\langle (|\mathbf{B}|-\langle |\mathbf{B}| \rangle)^2\rangle}^{1/2}/\langle |\mathbf{B}| \rangle$, and the normalized rms values of density fluctuations $\langle \delta n^2 \rangle^{1/2}/\langle n \rangle = {\langle(n - \langle n \rangle)^2} \rangle ^{1/2}/{\langle n \rangle}$. We also quantified the relationship between compressible fluctuations and the proton plasma beta, $\beta = 2\mu_0 n k_B T_p/|\mathbf{B}|^2$, with $k_B$ the Boltzmann constant, $T_p$ the proton temperature, and $\mu_0$ the vacuum permeability. 

\begin{table*}[t!]
\centering
\footnotesize
\renewcommand{\arraystretch}{1.05}
\setlength{\tabcolsep}{2pt}
\begin{tabular*}{\textwidth}{@{\extracolsep{\fill}}lccccccccc}
\hline
\multicolumn{9}{c}{\textbf{Alfvénic}} \\
\hline
Mission 
& $\langle |\mathbf{B}| \rangle$
& $| \langle \mathbf{B} \rangle|$
& $\langle|\delta {\bf B}_\perp|^2\rangle^{1/2}$
& $\langle|\delta {\bf B}_\parallel|^2\rangle^{1/2}$ 
& $\langle |\delta \mathbf{B}|^2 \rangle^{1/2}$
& $\frac{{\langle (\delta |\mathbf{B}|^2)^2 \rangle}^{1/2}}{\langle {|\mathbf{B}|^2}\rangle}$
& $\langle n \rangle$ 
& ${\langle (\delta n)^2 \rangle}^{1/2}$ 
& $\frac{{\langle (\delta n)^2 \rangle}^{1/2}}{\langle n\rangle}$\\
\hline
PSP & $-2.04 \pm 0.02$ & $-2.08 \pm 0.02$ & $-1.84 \pm 0.04$ & $-1.58 \pm 0.06$ & $-1.82 \pm 0.04$ & $0.86 \pm 0.06$ & $-1.75 \pm 0.07$ & $-1.97 \pm 0.06$ & $-0.28 \pm 0.06$  \\
SolO & $-1.55 \pm 0.01$ & $-1.56 \pm 0.01$ & $-1.60 \pm 0.02$ & $-1.61 \pm 0.03$ & $-1.60 \pm 0.02$ & $0.22 \pm 0.03$ & $-2.12 \pm 0.02$ & $-2.29 \pm 0.03$ & $-0.17 \pm 0.02$   \\
Combined & $-1.66 \pm 0.01$ & $-1.69 \pm 0.01$ & $-1.55 \pm 0.01$ & $-1.45 \pm 0.01$ & $-1.54 \pm 0.01$ & $0.43 \pm 0.01$ & $-2.11 \pm 0.01$ & $-2.52 \pm 0.01$ & $-0.44 \pm 0.01$   \\

\hline
\multicolumn{9}{c}{\textbf{Non-Alfvénic}} \\
\hline
PSP & $-1.79 \pm 0.29$ & $-1.78 \pm 0.51$ & $-1.18 \pm 0.32$ & $-1.23 \pm 0.50$ & $-1.20 \pm 0.34$ & $0.32 \pm 0.58$ & $-2.63 \pm 0.47$ & $-2.40 \pm 0.46$ & $0.25 \pm 0.54$   \\
SolO & $-1.62 \pm 0.13$ & $-1.56 \pm 0.15$ & $-1.65 \pm 0.19$ & $-2.02 \pm 0.26$ & $-1.75 \pm 0.20$ & $-0.35 \pm 0.18$ & $-2.12 \pm 0.17$ & $-2.38 \pm 0.24$ & $-0.25 \pm 0.17$   \\
Combined & $-1.45 \pm 0.06$ & $-1.42 \pm 0.07$ & $-1.38 \pm 0.08$ & $-1.51 \pm 0.11$ & $-1.41 \pm 0.08$ & $-0.08 \pm 0.08$ & $-2.02 \pm 0.08$ & $-2.09 \pm 0.10$ & $-0.07 \pm 0.07$ \\

\hline
\end{tabular*}
\caption{Radial dependence of the mean and rms of different quantities. Power law scalings are given separately for each mission and for the combined dataset, for both Alfvénic and non-Alfvénic solar intervals.}
\label{tab:radial_scalings}
\end{table*}

{In addition, we computed the sampling angle between wind velocity and the (scale-dependent) mean magnetic field 
$${\theta_{v B_0}} = \cos^{-1}\left(\frac{\mathbf{V}_{SW} \cdot \langle \mathbf{B} \rangle}{|\langle {\mathbf{B} \rangle||\mathbf{V}_{SW}|}}\right).$$
Under Taylor’s hypothesis, this angle quantifies the direction along which the fluctuations are advected past the spacecraft relative to $\langle{\bf B} \rangle$, and it does not necessarily coincide with the true propagation direction angle to the mean field. The statistics on the sampling angle (not shown) indicate that Alfvénic intervals are typically observed in field-aligned wind, with the most probable values of \({\theta_{v B_0}}\) around \(30^\circ\) and \(150^\circ\) for both the Wind and SolO datasets. In contrast, non-Alfvénic wind is found at very oblique angles, approximately \({\theta_{v B_0}} \sim 90^\circ\). Instead, PSP has almost consistently detected field-aligned wind for both types of winds. }
 
To identify fluctuations of the fast and slow mode type, we used the shorter sub-interval dataset to quantify the correlation between normalized magnetic pressure fluctuations ($\delta|\mathbf{B}|^2/\langle |\mathbf{B}|^2 \rangle={\langle (|\mathbf{B}|^2-\langle |\mathbf{B}|^2 \rangle)\rangle}/\langle |\mathbf{B}|^2 \rangle $) and density fluctuations $\delta n/\langle n \rangle$, {and we use the zero-lag cross-correlation between the two time-series 
$$
\text{Corr}(\delta n, \delta |{\bf B}|^2) =
\frac{\sum_{i=0}^{N-1}\delta n_i \  \delta |{\bf B}|_i^2}
{\sqrt{\sum_{i=0}^{N-1}\delta n_i^2}\sqrt{\sum_{i=0}^{N-1}{(\delta |{\bf B}|_i^2)}^2}}
$$}

Although such correlations should be calculated with  fluctuations of thermal pressure, the second order moment measurements are {inherently more sensitive to instrumental error and noise level than the zeroth and first order moments}. To mitigate this issue, we assumed isothermal conditions and considered fluctuations in density instead.

We used wavelet analysis to compute the coherence between $\delta|\mathbf{B}|^2/\langle |\mathbf{B}|^2 \rangle$ and $\delta n/\langle n \rangle$ \citep{torrence1998practical,torrence1999interdecadal,lion2016coherent}, 
$$C(t,\omega)_{\delta|\mathbf{B}|^2/,\delta n}=  S[ W_{\delta|\mathbf{B}|^2}^\ast(t,\omega) \, W_{\delta n}(t,\omega)],$$ 
and the phase 
$$\phi(t,\omega) =\tan^{-1} \left(\frac{\mathrm{Im} [C(t,\omega)_{\delta|{\bf B}|^2,\delta n}]}{\mathrm{Re}[C(t,\omega)_{\delta|\mathbf{B}|^2, \delta n}]}\right).$$ 
Here, $W(t,\omega)$ corresponds to the Morlet continuous wavelet transform, and $t$ and $\omega$ denote time and the frequency, with the relation $\omega \propto 1/a$ where $a$ is the wavelet scale. The operator $S$ is a smoothing function applied in both the time and frequency domains.

{The wavelet analysis describes the phase difference between two signals as a function of both time and scale (frequency), indicating whether the signals oscillate in phase or in anti-phase. Coherence, on the other hand, measures how strongly the phases of two signals are linked at a given time and frequency, serving as an indicator of localized correlation in the time–frequency. In our study, we apply wavelet analysis specifically to determine coherence and phase at large scales. Thus, from the time–frequency coherence and phase spectrograms, we calculate the mean values of \(C(t,\omega)\) and \(\phi(t,\omega)\) over the frequency range of interest, \(1/5 \le \omega \tau_c \le 1\), where \(\tau_c\) denotes the correlation time of the interval under consideration.}

\section{results}
\label{res}

\subsection{Radial scalings and scale dependence}
\label{rad_scalings}
\begin{figure*}[ht!]
\includegraphics[width=0.95\textwidth]{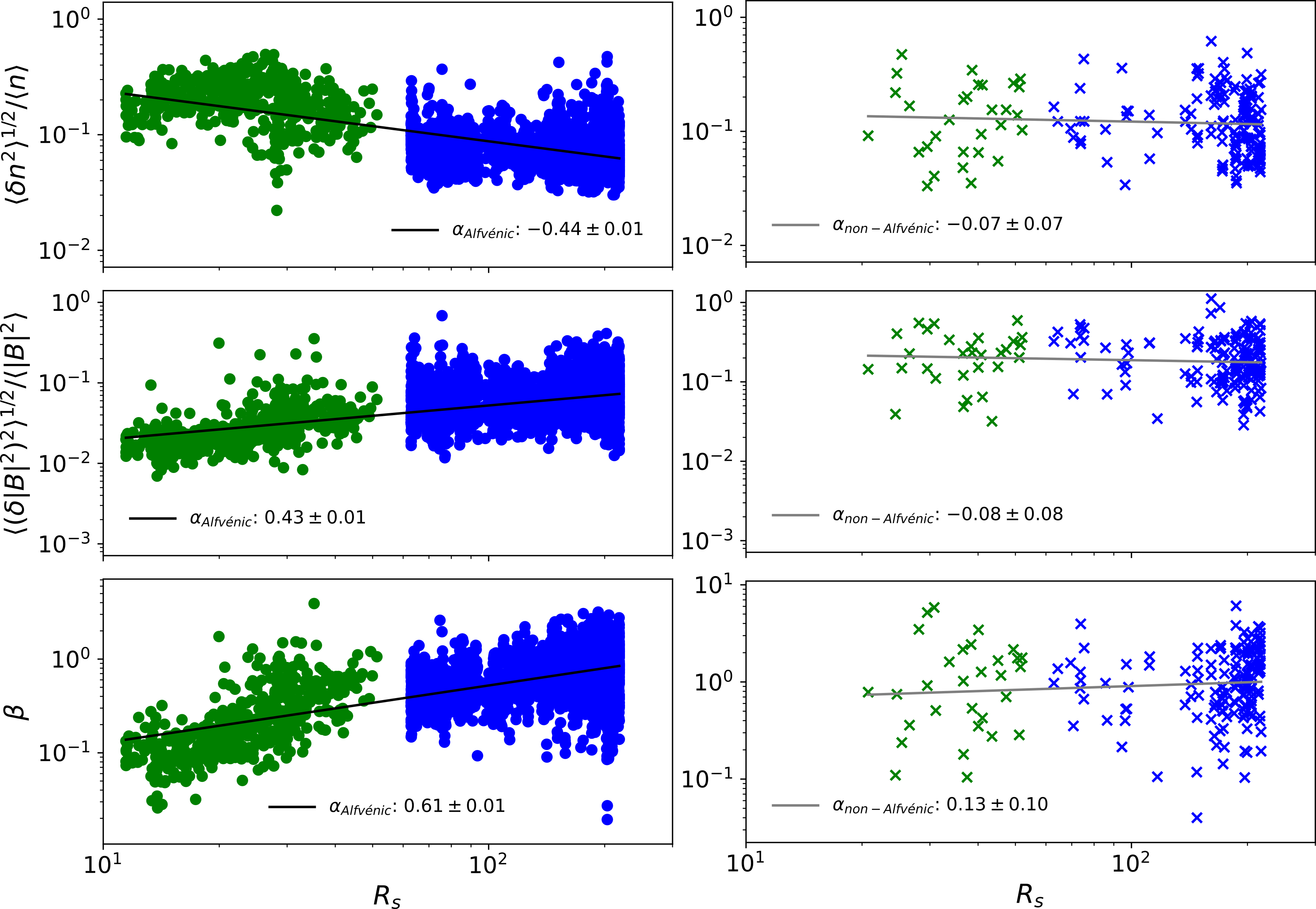}
  \caption{Values of normalized rms of the density (top panel) and magnetic pressure fluctuations (middle panel) and proton beta (bottom panel) averaged over each short sub-interval as a function of radial distance for Alfvénic (left column) and non-Alfvénic wind (right column). The  green and blue color correspond to PSP and SolO datasets, respectively. Alfvénic sub-intervals are marked with circles and  non-Alfvénic intervals with crosses. The power law coefficients from the best-fit curves are reported in the legend. The corresponding linear regression lines are plotted in black for the Alfvénic intervals and in gray for the non-Alfvénic intervals.}
\label{fig:one}
\end{figure*}

As expected from the overall expansion, both the mean magnetic field and the proton density decrease with radial distance. Table~\ref{tab:radial_scalings} reports the best fit parameters for several quantities, assuming a power-law dependence on the radial distance, $R^\alpha$. We list the resulting power-law index $\alpha$ for each individual dataset and for the combined set of observations. 
For the fitting procedure, we restricted the analysis to the short sub-intervals of the PSP and SolO datasets and calculated fluctuations and averages over a time scale $\tau=5\tau_c$ (corresponding to roughly $\tau \sim 0.68 \pm 0.44$ hours depending on radial distance). 

{When separating magnetic field into the mean and the fluctuating part $\mathbf{B} = \langle \mathbf{B} \rangle + \delta \mathbf{B}$, the WKB theory predicts  $\langle|\delta {\bf B}|^2\rangle^{1/2} \propto R^{-1.5}$, while the radial mean magnetic field $|\langle{\bf B}\rangle|\sim R^{-2}$.} The scaling of $|\langle {\bf B}\rangle |$ is consistent with a predominantly radial mean magnetic field during Alfvénic intervals in PSP, but including SolO measurements at larger distances yields a radial dependence of $R^{-1.66}$. On the other hand, the non-Alfvénic wind shows a slower decay of $R^{-1.45}$. These deviations from a purely radially expanding flux tube are likely associated with the Parker spiral geometry and possibly other large-scale structures.

For the rms magnetic fluctuation amplitude, we obtain $\langle|\delta {\bf B}|^2\rangle^{1/2} \propto R^{-1.54}$ and $\langle|\delta {\bf B}|^2\rangle^{1/2} \propto R^{-1.41}$ for Alfvénic and non-Alfvénic wind, respectively, in good agreement with earlier \citep{bavassano1982radial,tenerani2021evolution} and close to the WKB prediction (the faster decay is typically observed as the frequency of fluctuations increases). {However, if we decompose the fluctuations into components perpendicular and parallel to the background field, with the parallel component defined as $\delta \mathbf{B}_\parallel = \delta \mathbf{B} \cdot \hat{\mathbf{B}}_0$ where $\hat{\mathbf{B}}_0 = \langle \mathbf{B} \rangle/|\langle \mathbf{B} \rangle|$, and the perpendicular component $|\delta \mathbf{B}_\perp| = (\delta \mathbf{B}^2 - \delta \mathbf{B}_\parallel^2)^{1/2}$, we obtain $\langle|\delta {\bf B}_\perp|^2\rangle^{1/2} \propto R^{-1.55}$ and $R^{-1.38}$, and $\langle|\delta {\bf B}_\parallel|^2\rangle^{1/2} \propto R^{-1.45}$ and $R^{-1.51}$ for Alfvénic and non-Alfvénic solar wind, respectively.} In the Alfvénic wind (top three rows in Table \ref{tab:radial_scalings}), which is generally expected to follow WKB trends at large scales, the perpendicular component decays approximately as predicted (third column in Table \ref{tab:radial_scalings}), whereas the parallel fluctuating component displays a slower decay rate observed for the combined data (fourth column in Table \ref{tab:radial_scalings}). This slower radial decay of field‑aligned fluctuations with respect to the perpendicular ones is expected for incompressible, spherically polarized Alfvénic fluctuations with an exactly constant $B^2$, as a result of their coupling to the guiding magnetic field \citep{matteini2024alfvenic} --- although the specific decay rate of $\langle|\delta {\bf B}_\parallel|^2\rangle^{1/2}$ in general depends on the geometry of the mean field and can be affected by Parker's spiral effects. Nevertheless, the observed deviations from WKB of the field-aligned fluctuations could also indicate the generation of additional (subdominant) compressible fluctuations during the expansion. The latter interpretation is consistent with the observed increase in relative magnetic‑pressure fluctuations, which grow more rapidly in PSP and in the combined data set than in SolO (sixth column in Table \ref{tab:radial_scalings}), reflecting the same trend of $\langle|\delta {\bf B}_\parallel|^2\rangle^{1/2}$. 
In contrast, the radial evolution of the density in the Alfvénic wind is close to that expected from mass-flux conservation, with a scaling $\propto R^{-2.11}$, while the non-Alfvénic wind shows $R^{-2.02}$. The overall fitting of different plasma quantities are consistent with previous studies \citep{maruca2023trans}.

Despite the radial regression trends discussed above, fluctuations related to compressibility display substantial variability from one stream to another and are significantly scattered as a function of radial distance. Figure~\ref{fig:one} presents the normalized rms of the proton density and magnetic pressure fluctuations as a function of radial distance, while the bottom panel shows the mean plasma beta for each sub-interval. Although fitting the combined data indicates some dependence on radial distance, data points are highly dispersed around the regression line and only weakly correlated with $R$, with correlation coefficient $\text{corr}<0.5$ in all cases.

\begin{figure}[h!]
\includegraphics[width=0.49\textwidth]{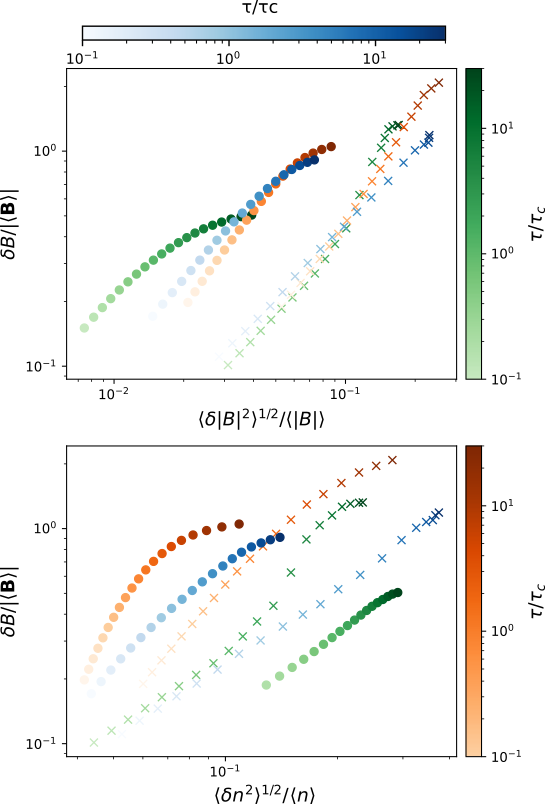}
  \caption{Scatter plots showing the ensemble-averaged distribution of the normalized rms of magnetic field fluctuations amplitude vs the ensemble-averaged normalized rms of density (top panel) and magnetic field magnitude (bottom panel) fluctuations obtained from the long-interval datasets.  Circles and crosses denote Alfvénic and non‑Alfvénic solar wind streams, respectively. The time scale $\tau$ is color coded and distinct colormaps are used to indicate PSP (green), SolO (blue) and Wind (orange) datasets.}
\label{fig:three}
\end{figure}

\begin{figure*}
\includegraphics[width=0.95\textwidth]{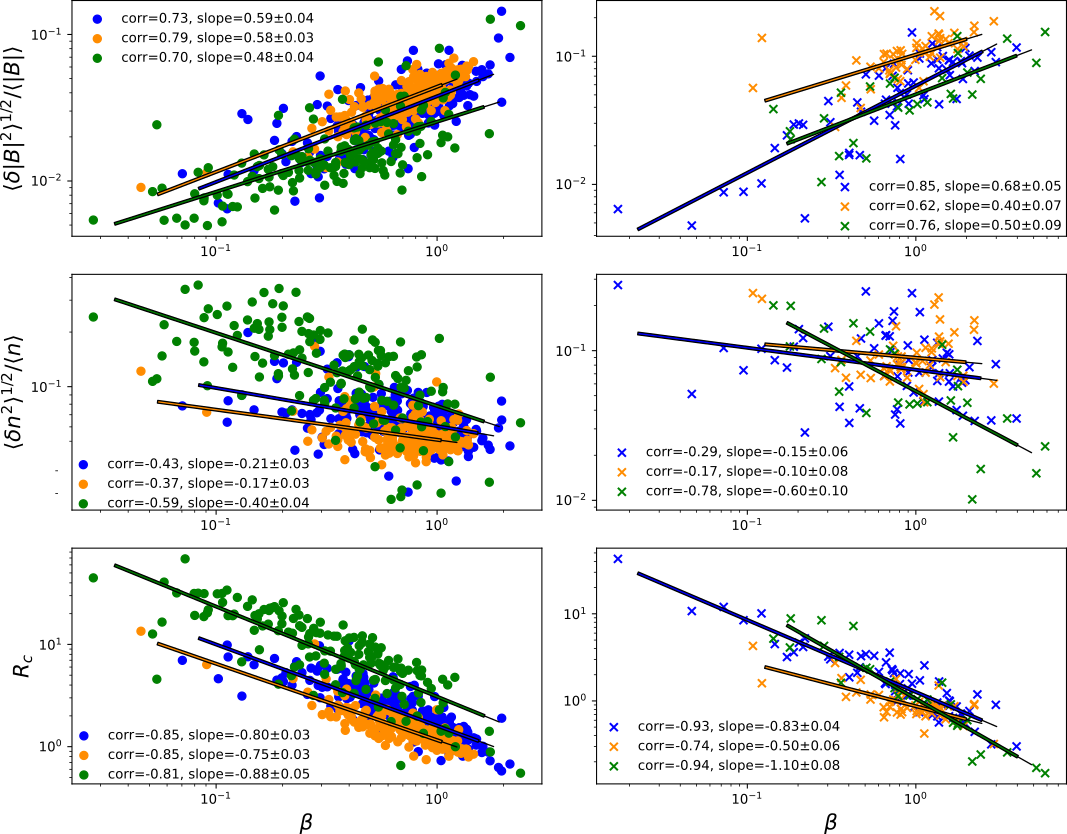}
  \caption{Correlation between compressible fluctuations and plasma beta. The left column shows results for Alfvénic intervals, while the right column corresponds to non-Alfvénic intervals. Magnetic field magnitude fluctuations are shown in the top panels, density fluctuations in the middle panels, and their ratio in the bottom panels. The correlation coefficient and the slope for each dataset is reported in the legend. Thick lines represent the linear regression fit for each mission, reported here as a reference for general trends.}
\label{fig:four}
\end{figure*}

We analyzed the scale-dependent properties of fluctuations indicative of compressibility within the inertial range, corresponding to scales $0.1<\tau/\tau_c<30$, using the long-interval dataset. The ensemble of time series of the magnetic field, magnetic field magnitude, and density fluctuation amplitudes were used to construct the probability density functions (PDFs) shown in Fig.~\ref{fig:appe1a}. The mean values of these PDFs are reported in Fig.~\ref{fig:three}, where different color shades represent increasing scale $\tau$ from light to dark color. Figure~\ref{fig:three} displays scatterplots of the rms of magnetic field fluctuations versus the rms of the normalized magnetic field magnitude fluctuations (top panel) and versus the rms density fluctuations (bottom panel). Circles represent Alfvénic intervals, while crosses  non-Alfvénic wind.

Alfvénic intervals display a characteristic saturation in amplitude at the largest scales, suggesting a possible threshold for relative fluctuation growth. This saturation is consistently observed across all datasets. This saturation of amplitudes is expected in the presence of spherically polarized fluctuations with constant magnetic field magnitude \citep{matteini2015ion,matteini20181,dunn2023effect}. Indeed, Alfvénic wind is generally less magnetically compressible than non-Alfvénic wind, with relative $|\mathbf{B}|$ fluctuations remaining below 10\%, although they tend to increase with radial distance. 
Non-Alfvénic intervals also exhibit an apparent amplitude saturation ($\delta B / | \langle \mathbf{B} \rangle|  \approx 1$) but their distributions have long tails extending beyond $\delta B /| \langle \mathbf{B} \rangle |\gg 2$, contrary to Alfv\'enic wind (as also shown by \cite{matteini20181}). Overall, non-Alfvénic wind is statistically more compressible than Alfvénic wind (up to three times the average rms values), with no clear radial dependence.

In contrast to magnetic field strength fluctuations, relative density fluctuations are enhanced closer to the Sun for Alfv\'enic wind, reaching relative amplitudes up to 30\% at the largest scales ($30 \ \tau/\tau_c$) consistent with previous near-Sun observations \citep{zhao2021analysis}, and decreasing for the largest scales to less than 10\% at 1 au. Non-Alfvénic intervals instead exhibit similar levels of density fluctuations at the largest scales across radial distances. 

\begin{figure*}[ht!]
\includegraphics[width=1\textwidth]{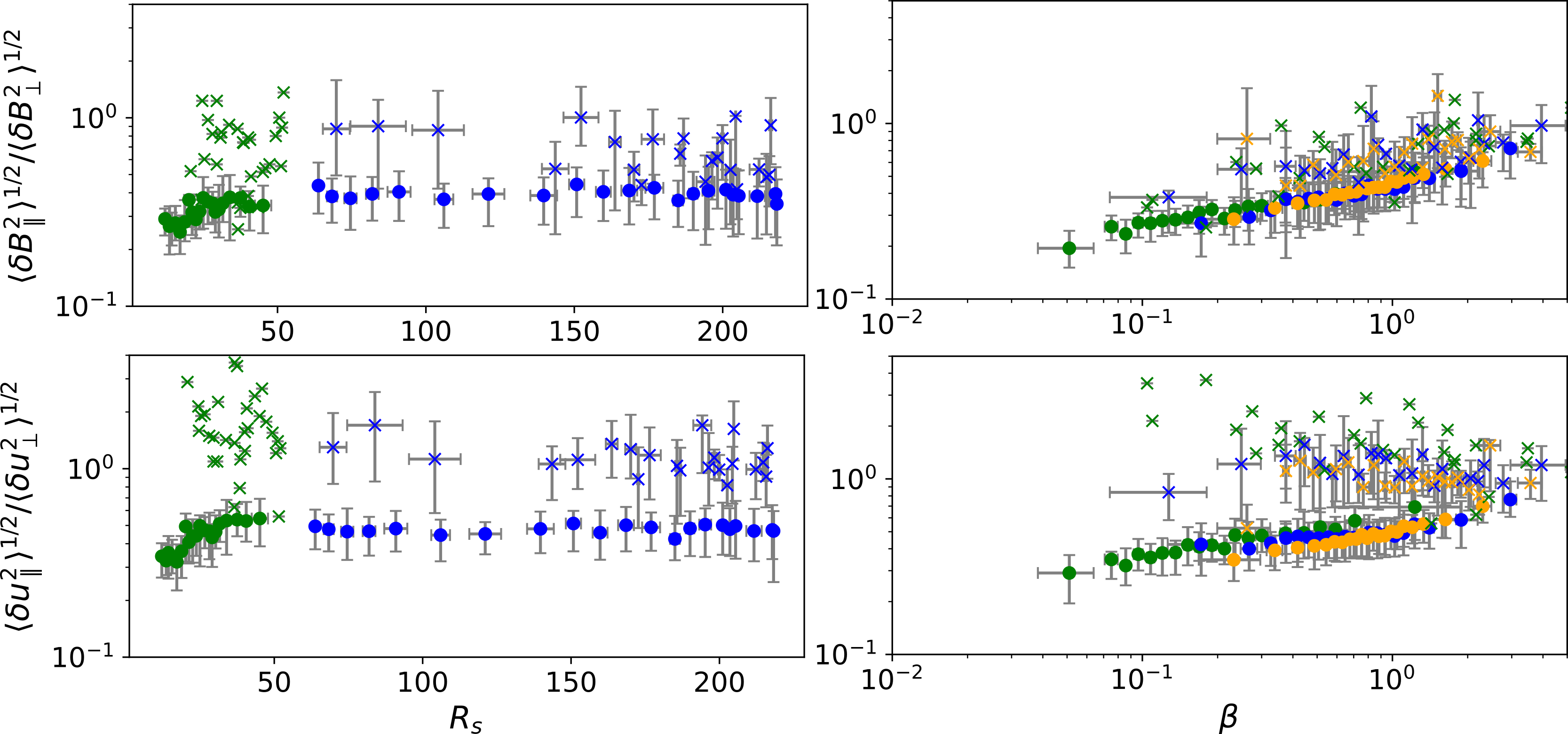}
  \caption{Ratios of the parallel to perpendicular component of magnetic field (top panels) and proton velocity (bottom panels) rms fluctuations as a function of radial distance (left) and plasma beta (right). Data points are obtained after averaging over each short sub-interval and binned, with bars indicating the variations of the data in each bin. Different colors are used to represent datasets from PSP (green), SolO (blue) and Wind (orange).}
\label{fig:seven}
\end{figure*}

\subsection{Dependence of compressible fluctuations on radial distance and plasma beta}

Although certain radial trends are observed in the above analysis, compressibility exhibit significant stream-to-stream variability. On the contrary, a stronger correlation with $\beta$ is observed for compressibility related fluctuations. Figure~\ref{fig:four} presents the normalized rms of the magnetic field magnitude (top panels) and of the density fluctuations (middle panels) computed for $\tau = 5\tau_c$. Each data point represents the  ensemble average over a long sub-interval and shown as a function of the average proton $\beta$ for each sub-interval. Similar results are found at individual time scales (not shown). We also plot the ratio $R_c=(\langle\delta n^2\rangle ^{1/2}/\langle n\rangle) /( \langle(\delta|{\bf B}|)^2\rangle^{1/2}/\langle|{\bf B}|\rangle)$ (bottom panels). A strong correlation is obtained between magnetic compressibility and $\beta$ across all datasets, with a correlation coefficient exceeding $\text{corr}>0.5$. This dependence is consistent with previous findings \citep{chen2020,brodiano2023statistical}. The density fluctuations also present a good correlation with  $\beta$ in the inner heliosphere, although the correlation weakens with increasing radial distance. This trend is evident for both Alfvénic and non-Alfvénic solar wind. Interestingly, the ratio $R_c$ displays a stronger correlation with $\beta$ than either measure alone, with  $\text{corr}>0.8$ in both Alfvénic and non-Alfvénic intervals. 

We investigated the relative importance of the field aligned component in the magnetic and velocity field fluctuations, often associated to compressible fluctuations,  relative to the transverse component, which contains mostly the Alfvénic contribution. Figure~\ref{fig:seven} shows  the ratio between the rms of the parallel and perpendicular components as a function of $R$ (left column) and $\beta$ (right column) calculated at the same scale $\tau=5\tau_c$ but by using the short sub-intervals to improve statistics. Instead of showing individual data points, the data were binned, and we show the mean value of each bin. The gray bars represent the spread of data within each bin, calculated using the 16th and 84th percentiles.

\begin{figure*}[ht!]
\includegraphics[width=1\textwidth]{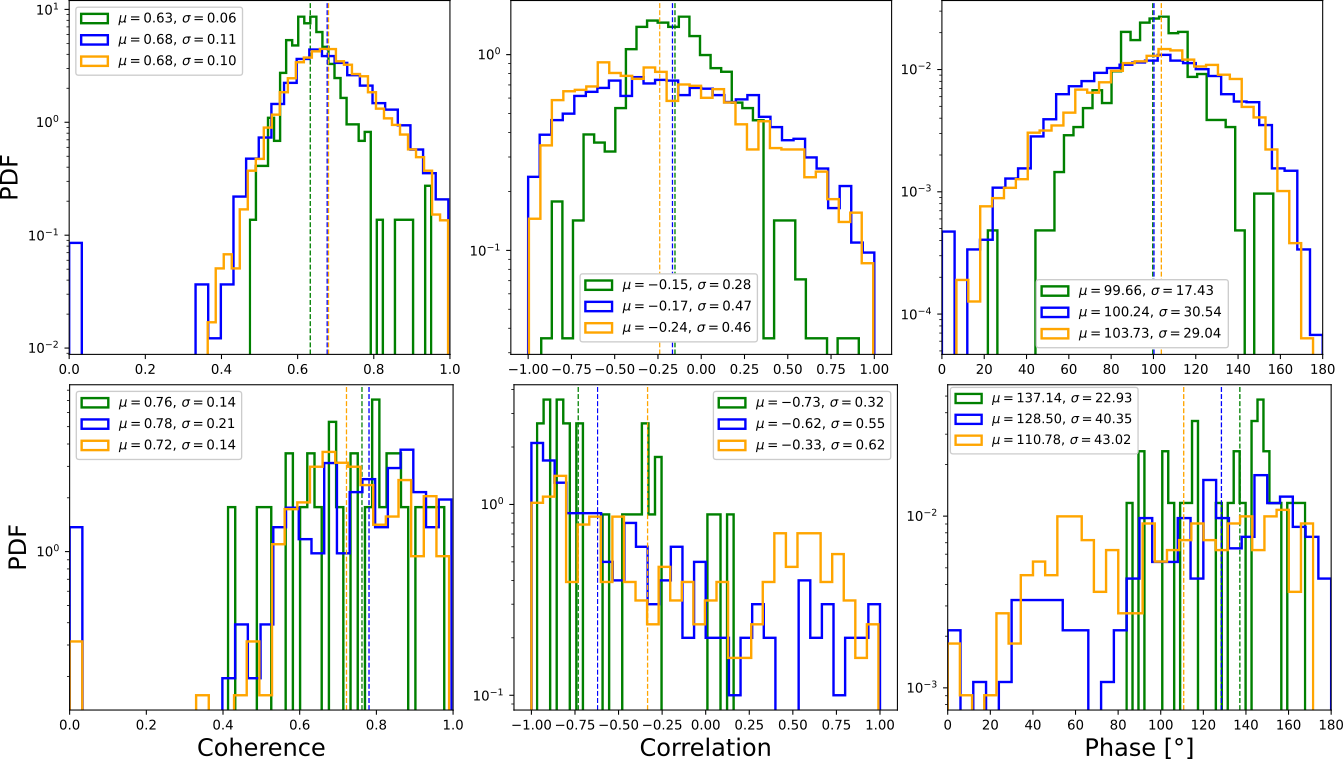}
  \caption{Distribution of correlation measures of compressible fluctuations for Alfvénic (top) and non-Alfvenic intervals (bottom). The panels show the PDFs of coherence (left), Pearson correlation (middle), and phase angle between $ \delta n/\langle n\rangle $ and $ \delta (|\mathbf{B}|^2)/\langle|\mathbf{B}|^2\rangle $ (right). Results are shown for Wind 
  (orange), SolO (blue), and for PSP (green). { The dashed lines represent the median for each distribution. The median and the standard deviation for each case are reported in the legend.}}
\label{fig:five}
\end{figure*}

For the Alfvénic wind, we observe that although the ratio remains significantly below unity, field-aligned fluctuations increase near the Sun compared to the transverse fluctuations up to $R\approx50 R_s$, and then their ratio saturates at a nearly constant value around 0.48 at larger distances. This behavior is consistent with the observed radial scalings laws of fluctuations already discussed in Section \ref{rad_scalings}. On the contrary, the ratio of field-aligned to transverse fluctuations in the non-Alfvénic wind does not show a well defined trend with radial distance and is characterized by large values of $\delta B_\parallel$ and $\delta v_\parallel$, which are comparable to, or even larger than, the perpendicular component. 

Interestingly, the previous observed dependence with $\beta$ for compressible fluctuations (Fig.~\ref{fig:four}) is also found here for the Alfv\'enic wind in both magnetic and velocity field data, and in the magnetic field data for the non Alfv\'enic wind. The plots on the right column show a clear monotonic dependence of the fluctuations ratios with $\beta$ that is distinct from the radial trends, and that qualitatively follows the trend of slow magnetosonic modes. 

This result further suggests that the evolution of field-aligned fluctuations in the magnetic field might be the result of the contribution from both expansion effects on spherically polarized states and local mechanisms controlled by the plasma beta.

\subsection{Statistical analysis of compressible fluctuations}
\label{sec:three}
\
To characterize compressible fluctuations, we computed the coherence and phase between $\delta(|\mathbf{B}|^2)/\langle |\mathbf{B}|^2 \rangle$ and $\delta n/\langle n \rangle$ using the wavelet analysis for each short sub-interval (see Sec. \ref{datasec}). In addition, we also quantified the linear correlation between magnetic and plasma compressible fluctuations by calculating the correlation coefficient. For this analysis, we considered fluctuations at the scale $\tau=5 \tau_c$, and the average coherence and phase between them are computed across the frequency range $1/5 < \omega {\tau_c} < 1$ and over each entire sub-interval. This methodology allows us to identify the coherence and phase at the scale of interest and determine whether the fluctuations are in or out of phase, providing insight into the polarization of compressible modes. For reference, Fig.~\ref{fig:appe*} shows an example of a wavelet analysis performed on an Alfvénic short sub-interval from the Wind database.

Figure~\ref{fig:five} shows the distribution of the correlation measures for both Alfvénic and non-Alfvénic intervals. Table~\ref{tab:table1} summarizes additional statistical information including the fraction of sub-intervals with strong positive or negative correlation ($|\text{corr}|>0.5$) and those with weak correlation ($|\text{corr}|<0.5$), filtered by $C>0.5$. The table also lists the mean values of the phase $\langle\phi^\pm\rangle$ and $\langle{\theta_{v B_0}^\pm}\rangle$ for each group. By eye inspection of some selected sub-intervals, we found that  low-correlation ($|\text{corr}|<0.5$) is often due to the simultaneous presence of both positive and negative correlated fluctuations within the 2-hour sub-interval considered that, when averaged over the length of the sub-interval, reduce the correlation. For this reason we believe that data with $|\text{corr}<0.5|$ still contain important information. Keeping these data is particularly relevant for the Alfv\'enic wind that shows a larger fraction of weakly correlated sub-intervals than the non-Alfv\'enic wind.

\begin{table}[t!]
\centering
\scriptsize
\renewcommand{\arraystretch}{1.0}
\begin{tabular}{@{}lrrrrrr@{}} 
\hline
\multicolumn{7}{c}{\textbf{Alfvénic}} \\
\hline
mission & $\text{portion}^+$  & $\text{portion}^-$ & $\langle  \phi^+ \rangle$ & $\langle \phi^- \rangle$ & $\langle  \theta_{v B_0}^+ \rangle$ & $\langle  \theta_{k B_0}^- \rangle$ \\
\hline
\multicolumn{7}{c}{${|\text{corr}|>0.5}$} \\
\hline
Wind & $10.15\%$ & $29.46\%$ & $49.61^\circ$ & $133.33^\circ$ & $50.2^\circ$ & $55.25^\circ$ \\
SolO & $11.19\%$ & $25.37\%$ & $50.33^\circ$ & $134.89^\circ$ & $49.46^\circ$ & $51.68^\circ$ \\
PSP  & $1.96\%$  & $8.70\%$ & $56.40^\circ$ & $128.74^\circ$ & $30.68^\circ$ & $31.78^\circ$ \\
\hline
\multicolumn{7}{c}{${|\text{corr}|<0.5}$} \\
\hline
Wind & $22.62\%$ & $35.66\%$ & $79.80^\circ$ & $105.35^\circ$ & $52.41^\circ$ & $53.84^\circ$ \\
SolO & $26.47\%$ & $33.89\%$ & $77.87^\circ$ & $105.67^\circ$ & $48.70^\circ$ & $50.21^\circ$ \\
PSP  & $26.09\%$ & $62.83\%$ & $82.53^\circ$ & $104.17^\circ$ & $33.90^\circ$ & $31.79^\circ$ \\
\hline
\multicolumn{7}{c}{\textbf{Non-Alfvénic}} \\
\hline
\multicolumn{7}{c}{${|\text{corr}|>0.5}$} \\
\hline
Wind & $22.28\%$ & $42.49\%$ & $48.61^\circ$ & $143.46^\circ$ & $65.17^\circ$ & $54.55^\circ$ \\
SolO & $10.39\%$ & $53.90\%$ & $45.79^\circ$ & $147.39^\circ$ & $64.53^\circ$ & $62.61^\circ$ \\
PSP  & $0.0$ & $60.00\%$ & - & $147.75^\circ$ & - & $43.11^\circ$ \\
\hline
\multicolumn{7}{c}{${|\text{corr}|<0.5}$} \\
\hline
Wind & $15.03\%$ & $17.10\%$ & $78.46^\circ$ & $109.47^\circ$ & $61.31^\circ$ & $53.84^\circ$ \\
SolO & $6.49\%$  & $21.43\%$ & $86.61^\circ$ & $112.33^\circ$ & $61.97^\circ$ & $55.25^\circ$ \\
PSP  & $10.00\%$ & $26.67\%$ & $89.17^\circ$ & $113.27^\circ$ & $42.27^\circ$ & $31.79^\circ$ \\
\hline
\end{tabular}
\caption{Summary of statistical results grouped by correlation sign in the Alfvénic and non‑Alfvénic wind datasets. The table reports, for each type of wind, the fraction of intervals with positive and negative correlation, together with the ensemble‑averaged phase and angle between the mean magnetic field and the solar wind velocity. Results are presented for both highly correlated and weakly correlated periods within each wind type.}
\label{tab:table1}
\end{table}

This statistical analysis shows that, at the scales considered, the compressible fluctuations are generally highly coherent across all datasets in both turbulent regimes. Within the non-Alfvénic intervals, we find similar statistical behavior at different heliocentric distances. Most of these intervals exhibit strong anti-correlation (with phase angles greater than $140^\circ$), characteristic of slow-mode fluctuations. A smaller fraction of intervals displays positive correlation and is nearly in phase, consistent with fast mode-like fluctuations.

By examining only intervals with strong correlations $(|\text{corr}|>0.5)$, it is evident that most of the non-Alfvénic solar wind is primarily composed of negatively correlated intervals. In fact, at the scales considered here, no positively correlated intervals were identified in the PSP data, and only a minor contribution is seen farther out. This supports the conclusion that slow modes are the dominant compressible component in non-Alfvénic, balanced turbulent solar wind.

The Alfvénic solar wind exhibits characteristics similar to the non-Alfvénic intervals, with a larger fraction of data characterized by anti-correlated fluctuations than by positively correlated ones. Nonetheless, in most Alfvénic intervals fluctuations are either uncorrelated or only weakly correlated. Across all datasets (PSP, SolO and Wind with weak correlation), these intervals maintain an approximately mean phase of $\phi^+\approx78^\circ$ for positively and $\phi^-\approx105^\circ$ negatively correlated cases.

In general, when aggregating both strong and weak correlations, we find that the Alfvénic wind exhibits $\approx30\%$ fast‑type fluctuations, compared to $\approx 20\%$ in the non-Alfv\'enic wind, and more than $\approx60\%$ of the fluctuations are of the slow‑type in both winds, reaching $>70\%$ in the slow wind. This may indicate that imbalanced turbulence systematically favors the generation of fast‑type modes compared with balanced turbulence, although the slow mode is dominant in both winds. {Figure~\ref{fig:append2} presents the correlation analysis for Wind dataset considering all sub-intervals and a reduced subset that includes only intervals where the relative difference between the two instruments is less than 10\%. The results indicate that slow modes are prevalent in both solar wind regimes and confirm that, even in this reduced sample, compressible fluctuations remain strongly anti-correlated, in agreement with prior results.}

\begin{figure*}[ht!]
\centering
\includegraphics[width=0.99\textwidth]{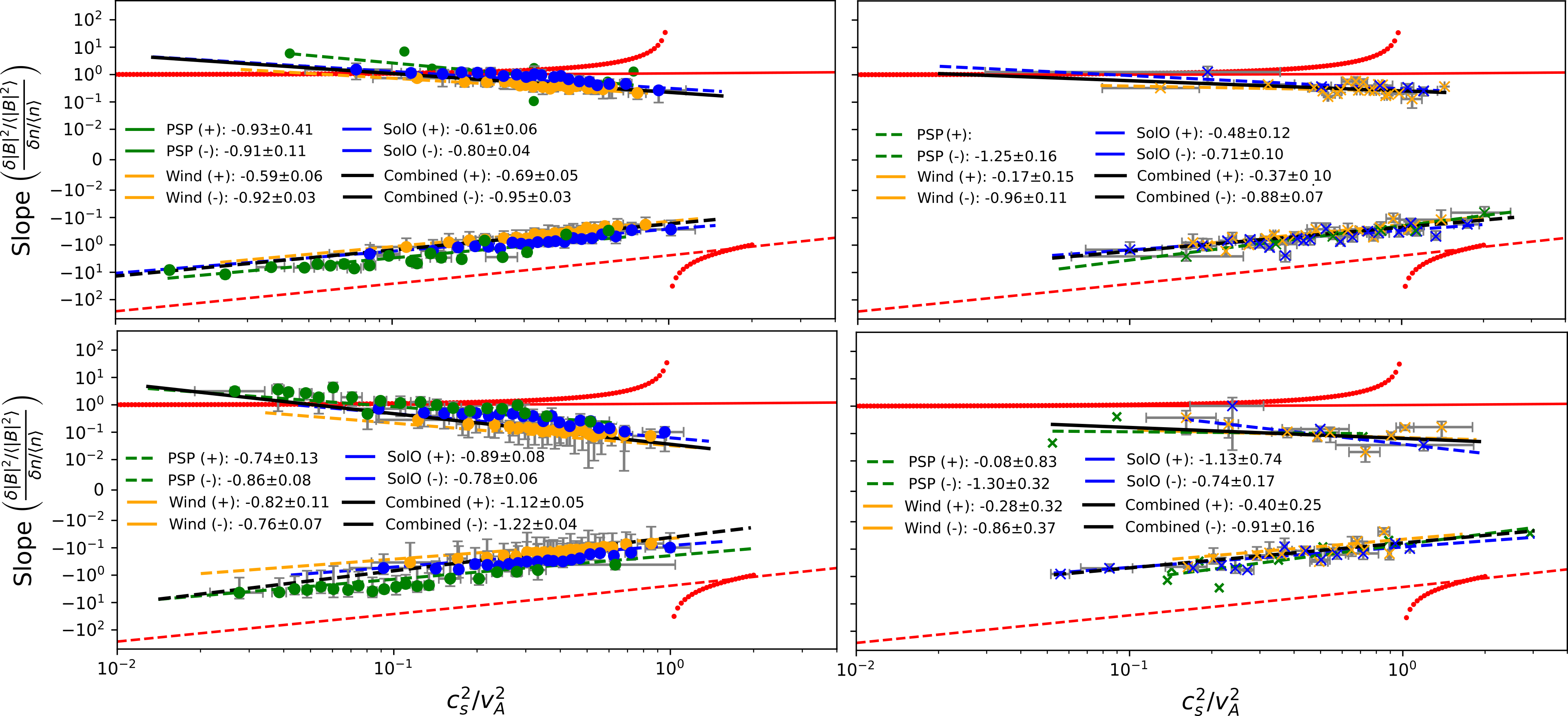}
  \caption{The response of density fluctuations to magnetic pressure variations in Alfvénic wind (left) and non-Alfvénic wind (right). The top panel correspond to intervals with high correlation  ($\text{corr}>0.5$), while the bottom panel show those with low correlation ($\text{corr}<0.5$). The error bars shows the standard deviations in the slope and $c_s^2/v_A^2$ values represented by each bin. The best fit of the combined data with positive and negative slopes is shown using the black lineschris. We also plot the expected response from fast (red solid) and slow (red dashed) linear MHD modes for waves propagating at $60^\circ$. The response prediction from forced modes are shown using red dots.}
\label{fig:six}
\end{figure*}

To explore the source of compressibility, we analyzed the response of density fluctuations to variations in magnetic pressure. This is quantified by creating a scatter plot of $\delta n/\langle n \rangle$ against $\delta (|\mathbf{B}|^2)/\langle |\mathbf{B}|^2 \rangle$ for each short sub-interval, from which we derived the value of the slope from their best linear fit (see bottom-left panel in Fig~\ref{fig:appe*}). Sub-interval were then categorized based on whether they exhibited a positive or negative slope. Figure~\ref{fig:six} displays the results for both Alfvénic and non-Alfvénic intervals with $|\text{corr}|>0.5$ (top panel) and $|\text{corr}|<0.5$ (bottom panel). We grouped the data into bins, representing values from all sub-intervals with similar slopes and beta values.

This analysis reveals a strong dependence on $\beta$, with low‑$\beta$ streams exhibiting a stronger response. This trend appears in both positively and negatively correlated periods and remains robust under both Alfvénic and non‑Alfvénic solar wind conditions. While magnetic compressibility is generally small in the near‑Sun, the low-$\beta$ values there allow for an enhanced response, which may explain the larger density fluctuations seen in PSP data than further out. We calculated the best fits for fast and slow mode-like fluctuations for each dataset and also the combined one, as indicated in the figure legends. For reference, we also display with red solid and dashed curves the linear MHD prediction for the relation between density and magnetic fluctuations for the fast ($+$) and slow ($-$) modes, 
\begin{equation}
{\delta n}/{\langle n\rangle} = {v_A^2}/{(v_{ph}^2 - c_s^2)}\,{\delta B_\parallel}/|\langle{\bf B}\rangle|,
\label{eq:mhd_corr}
\end{equation}
where the square phase speed is 
$$
v_{ph}^2= \frac{1}{2}\left[(v_A^2 + c_s^2) \pm \sqrt{(v_A^2 + c_s^2)^2 - 4 v_A^2 c_s^2 \cos^2(\theta_{v B_0})}\right],
$$
with $v_A$ and $c_s$ the Alfv\'en and sound speed and $c_s^2/v_a^2=\gamma\beta_T/2$ (where we fix $\gamma=1$ and $\beta_T$ is the total plasma beta that we approximate as $\beta_T\approx 2\beta$) \citep{schekochihin2022lectures,lee2020turbulence}. {Note that we calculate the proportionality coefficient in eq.~\eqref{eq:mhd_corr} between the density fluctuations and the nonlinear magnetic pressure fluctuations. The latter recovers the theoretical $\delta B_\parallel/|\langle{\bf B}\rangle|$ in the limit of small amplitude fluctuations. By inspection of the distributions in Fig.~\ref{fig:appe1a} and from Fig.~\ref{fig:three}, we see that this condition is not always satisfied, although the normalized mean values of both density and magnetic fluctuations are generally less than unity. Additionally, to make meaningful comparisons while maintaining the analysis simple, we used the value $\theta_{v B_0}=60^\circ$ (corresponding to the mean sampling angle calculated over all datasets considered) as reference value for the propagation angle $\theta_{kB_0}$}. We further show as red dots the expected trend of density fluctuations driven nonlinearly by an amplitude-modulated Alfvén wave predicted by MHD theory  \citep{cohen1974nonlinear,hollweg1971density,spangler1982properties} 

\begin{equation}
{\delta n}/{\langle n\rangle} = {(1 - c_s^2/v_A^2)}^{-1}\,{\delta (|\mathbf{B}|^2)}/|\langle{\bf B}\rangle|^2,
\end{equation}
{where we note that $\delta (|\mathbf{B}|^2)/\langle|{\bf B}|\rangle^2\simeq \delta (|\mathbf{B}|^2)/|\langle{\bf B}\rangle|^2$ by neglecting terms $\sim \langle|\delta {\bf B}|^2\rangle/|\langle{\bf B}\rangle|^2$.} 

We observe that intervals with negative correlation exhibit an approximate $\beta^{-1}$ scaling (shown as a red dashed line), in line with the predictions of MHD slow-mode theory \citep{verscharen2017kinetic}. This result implies that compressive fluctuations, both close to the Sun and at larger heliocentric distances, retain characteristics consistent with MHD-like slow modes. Since these intervals represent the majority of the dataset, we can argue that the slow mode is the most significant contributor to compressible fluctuations in the solar wind.

 In contrast, the positively correlated intervals do not display the MHD behavior anticipated from linear (shown as solid red line) or nonlinear (red dots) theory. The non-Alfv\'enic wind approaches the prediction of fast magnetosonic mode, with a shallower slope than $\beta^{-1}$, but still steeper than predictions. Instead, we find that in the Alfv\'enic wind fluctuations follow a nearly \(\beta^{-1}\) scaling, both for the full data set and for each mission considered separately. This feature is especially important for intervals with \(|\mathrm{corr}| < 0.5\), which constitute the majority of the positively correlated sample in the Alfv\'enic wind, as indicated in Table \ref{tab:table1}. Furthermore, we do not observe the predicted transition from positive to negative correlation at a specific critical $\beta$, as would be expected if the compressive response was driven by amplitude-modulated Alfvén waves. Thus, contrary to the slow-type fluctuations, at least at the level of our fits, the observations of fast-type correlated fluctuations do not match current theoretical expectations for fast MHD modes and we do not find trends that are consistent with MHD theory of forced compressible modes. Additional theoretical and numerical work, exploring different dimensionalities and geometries (including the effects of expansion), as well as extensions to kinetic theory is likely required to better understand the nature of the fast modes observed in the data.

\section{Summary and conclusions}
\label{conclusions}

In this work we studied the statistical properties of compressible fluctuations in the solar wind using \textit{in-situ} measurements obtained from PSP, SolO, and Wind missions, covering radial distances ranging from 10 solar radii to 1 au. We identified streams dominated by balanced and imbalanced turbulence in which the Alfvénic properties ($|\sigma_c| < 0.25$ and $|\sigma_c| > 0.75$) remained approximately constant for at least 30 correlation times. This methodology enabled us to analyze large-scale properties while reducing the mixing of streams from different origins and plasma conditions. We analyzed the radial scaling of magnetic field, magnetic field magnitude and density fluctuations, and compared with theoretical predictions of MHD waves. We also studied the scale dependence of the fluctuation amplitude and the level of compressibility at different radial distances and as a function of plasma beta. Finally, we performed a statistical analysis of the properties of compressible fluctuations in the solar wind.  

We found that magnetic compressibility is low near the sun and increases with heliocentric distance, whereas density fluctuations show the opposite trend, exhibiting enhanced plasma compressibility closer to the sun. By analyzing how these quantities correlate with radial distance and $\beta$, we argued that this behavior likely results from a combination of expansion effects and compressible dynamics governed by local plasma conditions. 
As the solar wind expands, relative density fluctuations are expected to decrease with distance or be reduced by kinetic damping. In contrast, small but finite magnetic field magnitude fluctuations are influenced by the radial increase of $\delta B/|\langle{\bf B}\rangle|$.  Together, these two effects may explain the opposite radial trends of plasma and magnetic compressibility.
However, the strong correlation of density and magnetic pressure fluctuations (Fig.~\ref{fig:four}), as well as of the ratio $\delta B_\parallel/\delta B_\perp$ (Fig.~\ref{fig:seven}), with plasma $\beta$ indicates that, in addition to the global radial evolution, local compressible processes controlled by $\beta$ also play an important role.

A statistical analysis of the correlation, coherence and phase between density and magnetic pressure fluctuations shows that both solar wind types exhibit highly coherent compressible fluctuations displaying positive or negative correlation (Fig.~\ref{fig:five} and Table~\ref{tab:table1}). Non-Alfvénic intervals are dominated by anti-correlated fluctuations, which account for more than 70\% of the short sub-intervals. In contrast, Alfvénic intervals contain a mixture of positively and negatively correlated compressible fluctuations, although the negatively correlated ones remain prevalent. Comparison between with MHD predictions for the relation between density and magnetic pressure fluctuations as a function of $\beta$ (Fig.~\ref{fig:six}) indicates that the dominant anti-correlated fluctuations are consistent with slow MHD magnetosonic modes, whereas the positively correlated fluctuations are not reproduced by either linear or nonlinear MHD models. Because slow modes also dominate in the Alfvénic wind, the observed trend of $\delta B_\parallel/\delta B_\perp$ versus $\beta$ (Fig.~\ref{fig:seven}, top right panel) is likewise consistent with slow‑mode magnetosonic behavior. Since $\beta$ is generally lower at PSP than at larger heliocentric distances, this dominance can also account for the enhanced density fluctuations observed in the Alfvénic wind by PSP, together with the radial increase in $\delta B_\parallel/\delta B_\perp$ and the comparatively faster growth of magnetic pressure fluctuations inside $50,R_s$ (Fig.~\ref{fig:seven}, top left panel; Table~\ref{tab:radial_scalings}).

{A considerable number of sub-intervals in the Alfvénic database exhibited low correlation because, within a single interval, shorter-scale phase variations with both positive and negative correlations were present. These sub-intervals nonetheless exhibit strong coherence, and their features resemble linear MHD modes, as shown in Figs.~\ref{fig:five} and \ref{fig:six}. In contrast, the positively correlated intervals deviate from MHD predictions and instead exhibit an approximate $\beta^{-1}$ dependence.}

{Our analysis concentrates on fluctuations within the inertial range, specifically at scales much larger than the characteristic proton kinetic scales—by three to four orders of magnitude compared with the proton gyroradius. This scale separation ensures that fluctuations fall within the MHD regime. However, as the turbulent cascade extends toward kinetic scales, kinetic modes (such as KAWs/KSWs) can arise and potentially contribute to compressive fluctuations. At low frequencies, KAWs and KSWs display similar observational characteristics, including anti-correlation between plasma and magnetic pressure, which makes them difficult to distinguish \citep{zhao2014properties}. Although the possible contribution of these modes in the observed compressibility cannot be completely ruled out, their study would require electric field measurements together with higher-cadence observations, in order to characterize polarization properties and its scale dependence down to kinetic scales, which lies beyond the scope of the present work.}

In conclusion, our analysis shows that density and magnetic pressure fluctuations are strongly influenced by plasma beta, demonstrating the presence of ongoing compressible dynamics within the expanding solar wind. Our results also support the idea that the enhanced density fluctuations observed near the Sun are produced by slow mode waves, which may contribute to solar wind heating and acceleration in that region \cite{kellogg2020heating,mozer2022core,kellogg2024heating}. A possible candidate for generating such fluctuations is parametric decay of Alfv\'en waves, which is expected to be favored in the low-$\beta$ regions close to the sun \citep{tenerani2017parametric,shoda2019three}. In the Alfvénic wind, fast mode-like waves also contribute to the compressible component. These positively correlated fluctuations in density and magnetic pressure could arise from several mechanisms, particularly those associated with forced modes generated by Alfvén-wave steepening, which are especially relevant in Alfvénic wind. However, we do not observe theoretical features such as the predicted transitions between fast and slow modes at critical plasma beta \citep{hollweg1971density,cohen1974nonlinear,mallet2021evolution}. These discrepancies motivate further comparisons with extended fluid and kinetic models, as well as numerical simulations.

\begin{acknowledgments}
This research was supported by NSF SHINE award number 2400967 and NSF CAREER award 2141564.  We acknowledge Wind, Solar Orbiter and Parker Solar Probe missions for the use of the data publicly available at \href{https://spdf.gsfc.nasa.gov/}{the NASA Space Physics Data Facility}. The CME catalog compiled by \cite{mostl2020prediction} is available at \href{https://helioforecast.space/icmecat}{https://helioforecast.space/icmecat}. CAG acknowledges Daniel Ipia-Achury, Trevor Bowen, Alfred Mallet, Ben Alterman, and Nahuel Andres for their valuable discussions regarding the results presented in this work. 

\end{acknowledgments}

\bibliographystyle{aasjournalv7}

\appendix
\label{appendix}
\setcounter{figure}{0}
\renewcommand{\thefigure}{A\arabic{figure}}
\section{Additional Figures}

{Figure~\ref{fig:appe1a} shows the probability distribution of $\delta B/|\langle \mathbf{B} \rangle|$, $\langle (\delta |{\bf B}|)^2 \rangle^{1/2}/\langle |{\bf B}| \rangle$ and $\langle \delta n^2\rangle^{1/2}/\langle n \rangle$ calculated at different scales $\tau/\tau_c$ (indicated in the colored top bars for each mission: green for PSP, blue for SolO and orange for Wind) across different solar wind regimes. For each distribution, the data are obtained by compiling time series of all quantities evaluated over multiple time scales for the long intervals of each dataset. The scale dependence is examined over the range $0.1<\tau/\tau_c<30$, and the mean value for each distribution is presented in Fig~\ref{fig:three}.}

\begin{figure*}[ht!]
\centering
\includegraphics[width=0.95\textwidth]{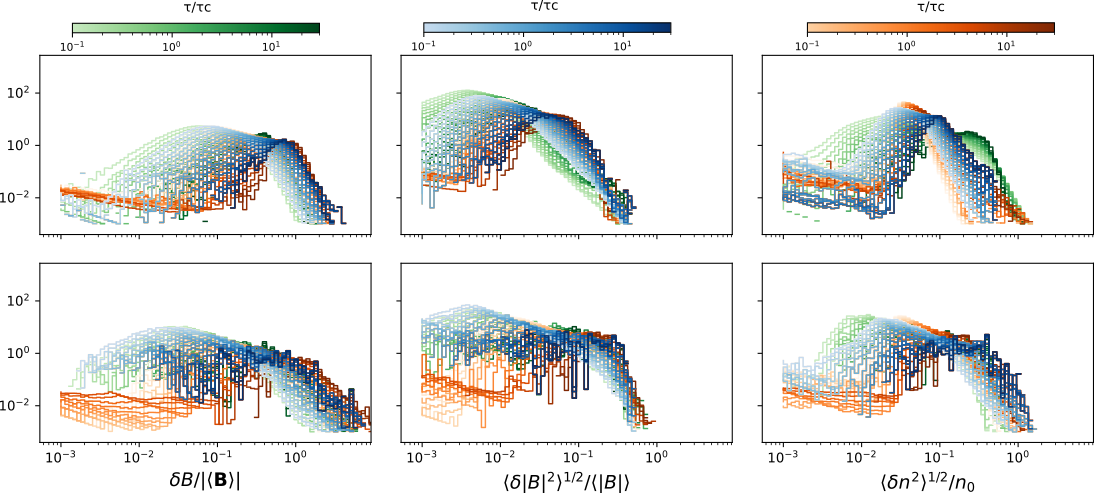}\\ 
  \caption{The PDF of the rms of the fluctuation's amplitude $\delta B/|\langle \bf{B} \rangle|$ (left), magnetic field magnitude fluctuations $\langle (\delta |{\bf B}|)^2 \rangle^{1/2}/\langle |{\bf B}| \rangle$ (middle), and of density fluctuations $\langle\delta n^2 \rangle^{1/2}/\langle n \rangle$ (right) for the Alfvénic (top panels) and non-Alfvenic wind (bottom panels). Different colors correspond to fluctuations over different time scales covering the range $0.1 <\tau/\tau_c<30$, as indicated in the top bars.}
\label{fig:appe1a}
\end{figure*}

{Figure~\ref{fig:appe*} provides an example interval used to illustrate the correlation and wavelet analysis described in Section~\ref{sec:three}. In the top-left panel, we plot the time series of the normalized density and magnetic pressure for this interval at the scale $\tau = 5\tau_c$. The bottom-left panel displays the scatter plot of $\delta n/\langle n \rangle$ versus $\delta |\mathbf{B}|^2/\langle |\mathbf{B}|^2 \rangle$. The legend reports the average proton beta for the interval together with the slope obtained from a linear regression. The panels on the right present the coherence and phase spectrograms calculated from these two time series. We then determined the mean coherence and phase within the frequency band of interest ($1/5 \le \omega \tau_c \le 1$) and over the interval. The resulting averages from all datasets are subsequently combined to obtain the statistics shown in Fig.~\ref{fig:five}. For the wavelet analysis we excluded regions of the time–scale space affected from edge effects near the start and end of the signals. Consequently, both the analysis and its interpretation are restricted to the portion lying inside the domain of influence}

\begin{figure*}[ht!]
\centering
\includegraphics[width=0.95\textwidth]{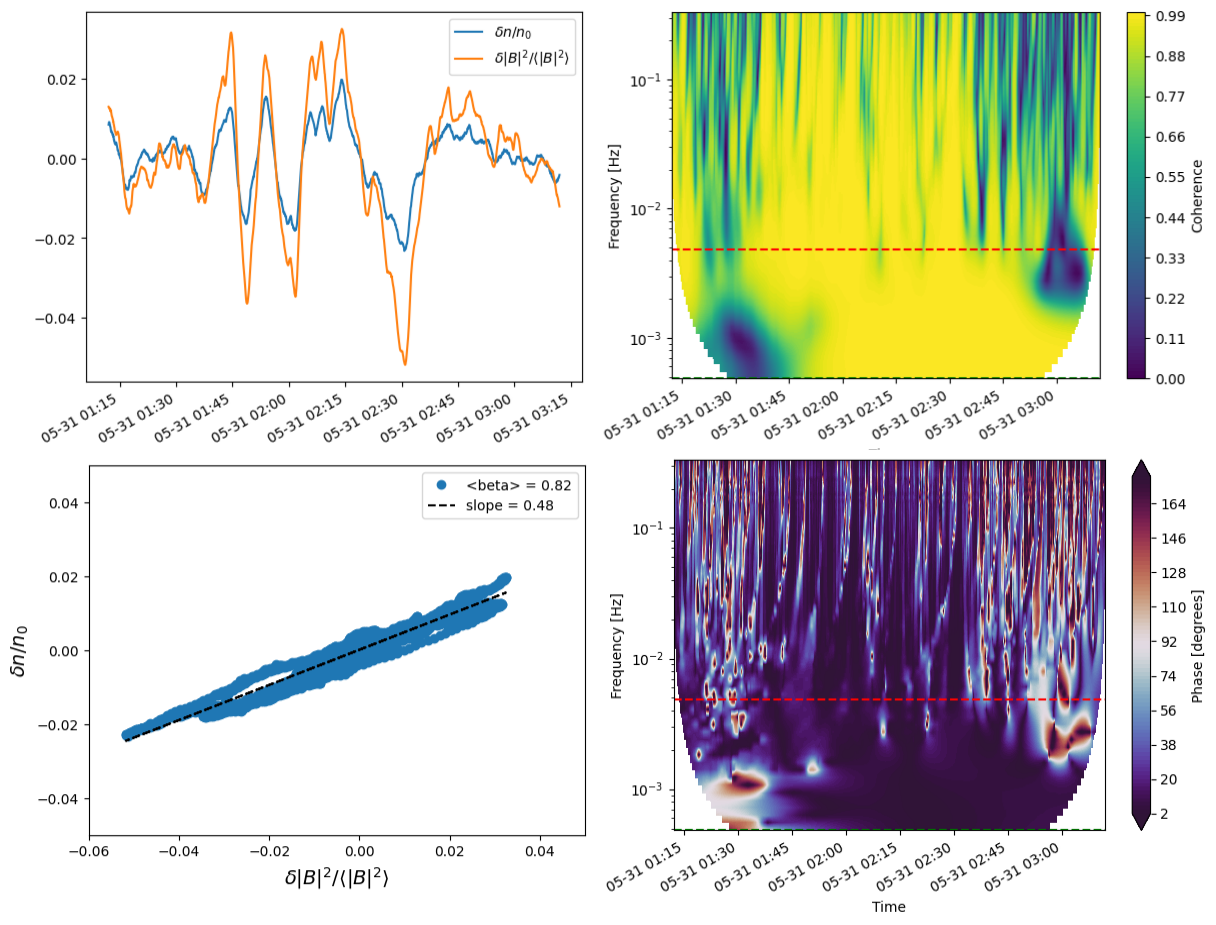}\\ 
  \caption{(Top Left) Time series of normalized density fluctuations ($\delta n/\langle n \rangle$) and magnetic pressure fluctuations ($\delta |\mathbf{B}|^2/\langle |\mathbf{B}| \rangle$) at the scale $\tau = 5 \tau_c$ for an Alfvénic short sub-interval of Wind observations spanning 2000-05-31 01:12:03 to 2000-05-31 03:11:45. (Bottom Left) Scatter plot of $\delta n/\langle n \rangle$ versus $\delta |\mathbf{B}|^2/\langle |\mathbf{B}| \rangle$, with the legend indicating the mean proton beta and the linear fit slope between these two time series. The top-right panel show the spectrograms of the coherence ($C(t,\omega)$) and the bottom-right the phase ($\phi(t,\omega)$) between density and magnetic pressure fluctuations. The red dashed line mark the correlation time for that interval.}
\label{fig:appe*}
\end{figure*}

\begin{figure}[ht!]
\begin{center}
\includegraphics[width=\textwidth]{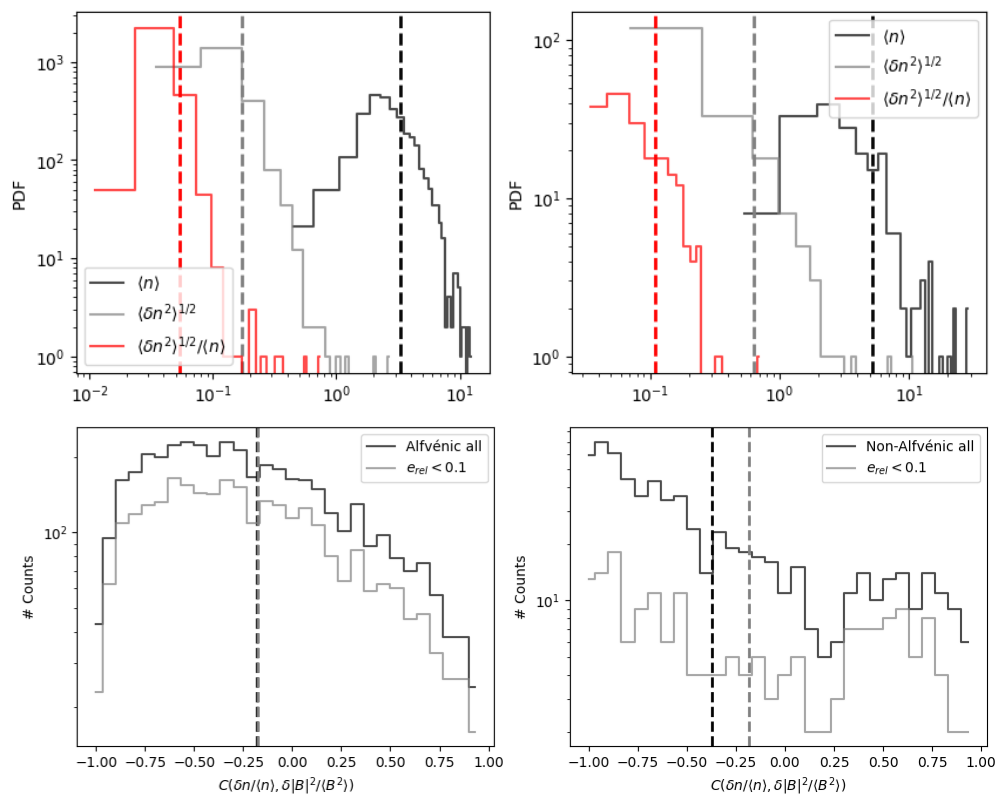}
\end{center}
\caption{(Top) PDF of the mean number density (black), the rms density fluctuations (gray), and the rms of the normalized density fluctuations (red) for Alfvénic intervals (left panel) and non-Alfvénic intervals (right panel) in the Wind dataset.
(Bottom) PDF of the correlation between normalized density and magnetic pressure fluctuations for each solar wind category, shown for the full dataset (black) and for the filtered dataset (gray).}
\label{fig:append2}
\end{figure}

{The top panel of Fig~\ref{fig:append2} presents the distribution of the mean density ($\langle n\rangle$), the rms density fluctuations ($\langle \delta n^2\rangle^{1/2}$), and the normalized density fluctuations ($\langle \delta n^2\rangle^{1/2}/\langle n\rangle$) at the scale $\tau = 5 \tau_c$ for Alfvénic (left) and non-Alfvénic (right) intervals in the Wind dataset. The bottom panel shows the PDF of the correlation between $\delta n/\langle n  \rangle$ and $\delta |\mathbf{B}|^2/\langle |\mathbf{B}|^2 \rangle$ for both the full and the filtered dataset ($\epsilon < 0.1$).}

\end{document}